\documentclass[traditabstract]{aa}
\usepackage{color}
\usepackage{comment} 
\usepackage{graphicx}
\usepackage{txfonts}
\usepackage{pstricks}
\include{epsf} 
\usepackage{natbib}
\usepackage{multirow}
\usepackage{color, colortbl}

\bibpunct{(}{)}{;}{a}{}{,}

\def\0{\phantom0}

\newcommand{\bold}[1]{ {\bf #1} }

\begin{document}

\title{Hopfield Neural Network deconvolution for \\ weak lensing measurement }

\titlerunning{Hopfield Neural Network deconvolution for weak lensing measurement }

\author{G. Nurbaeva\inst{1} \and M. Tewes\inst{2, 1} \and F. Courbin\inst{1} \and  G.~Meylan\inst{1} }

\institute{Laboratoire d'astrophysique, Ecole Polytechnique
     F\'ed\'erale de Lausanne (EPFL), Observatoire de Sauverny,
     CH-1290 Versoix, Switzerland
\and  Argelander-Institut f\"ur Astronomie, Auf dem H\"ugel 71, D-53121 Bonn, Germany \label{bonn}
  }

\date{Received; accepted }
 
\abstract{ Weak gravitational lensing has the potential to place tight constraints on the equation of the state of dark energy. However, this will only be possible if shear measurement methods can reach the required level of accuracy. We present a new method to measure the ellipticity of galaxies used in weak lensing surveys. The method makes use of direct deconvolution of the data by the total Point Spread Function (PSF). We adopt a linear algebra formalism that represents the PSF as a Toeplitz matrix. This allows us to solve the convolution equation by applying the Hopfield Neural Network iterative scheme. The ellipticity of galaxies in the deconvolved images are then measured using second order moments of the autocorrelation function of the images.
To our knowledge, it is the first time full  image deconvolution  is used to measure weak lensing shear. We apply our method to the simulated weak lensing data proposed in the GREAT10 challenge and obtain a quality factor of $Q=87$. This result is obtained after applying image denoising to the data, prior to the deconvolution. The additive and multiplicative biases on the shear power spectrum are then \mbox{$\sqrt{\mathcal{A}}=+0.09\times10^{-4}$} and \mbox{$\mathcal{M}/2=+0.0357$} respectively.} 


\maketitle
\section{Introduction}
\label{section:intro}

Weak gravitational lensing is one of the most powerful cosmological probes to study the nature of dark energy and its evolution with redshift. An important step towards achieving this goal relies on the accurate measurements of the so-called ``Cosmic Shear", i.e., the tiny change in ellipticity of the observed galaxies due to foreground cosmic structures. The measurement must be performed statistically, using millions of galaxies whose images are affected by the convolution of the atmospheric and instrumental Point Spread Function (PSF) and by noise. Reliable  image processing techniques are required to correct for the PSF smearing and to extract the faint Cosmic Shear signal.

So far, only a handful of truly different shape measurement algorithms are available. They split into four broad categories. The first known as ``forward fitting" methods fit analytical model of galaxies to measure their shape, decontaminated from the PSF \citep[e.g.][] {Gentile_gfit2012, Kacprzak2012, Miller2007}.  A second category decomposes the image of galaxies on a basis of functions of different shapes and reconstructs the original image \citep[e.g.,][]{Refregier2003, Kuijken2006}, again decontaminated from the PSF. The third class of methods is based on the calculation of the moments of the light distribution of galaxies and of their associated PSFs \citep{KSB1995, Bernstein2010}. More recently, methods based on machine learning have also started to be implemented \citep[e.g.][]{Gruen2010, Tewes2012}.

As the above methods show different levels of sensitivity to noise and to various sources of systematic errors, it is worth exploring further ways of performing PSF correction in weak lensing measurements. One way of doing this is to carry out full image deconvolution of the data prior to the shape measurement, which is the work undertaken in this article. Our new algorithm proposes to describe the image distortion by the PSF in a matrix equation using linear algebra formalism. In this formalism, the deconvolution problem can be solved by applying the Hopfield Neural Network iterative scheme. The galaxy ellipticities are then measured using the second-order moments of the two-dimensional auto-correlation function (ACF) of deconvolved images.

In Sect.~\ref{section:method}  we introduce the image deconvolution problem and discuss the method based on Hopfield Neural Networks to solve it. 
In Sect.~\ref{section:Galaxy shape measurement}   we present our new adaptive neuron updating rule for the astronomical image reconstruction and the  shape measurement algorithm for practical implementation.  Sect.~\ref{section:Parameter_lambda} deals with the calibration of the Tikhonov regularization parameter on simulated data in order to find the optimal value.
In  Sect.~\ref{section:application to the GREAT10 data} we present our application of the method to the data of the GREAT10 Galaxy challenge \citep{GREAT10Handbook2010, G10results}  and we  give a conclusion in Sect.~\ref{section:conclusion}.

\section{Image Deconvolution problem}
\label{section:method}

An observed  image can be modeled as a function $A$ of the true source image augmented by a noise term:

\begin{equation}
u_{0} = A(u_s) + n,
\label{Equ0_1}
\end{equation}
where $u_{0} $ and $u_s$ denote the observed image and the source image, respectively, and $n$ is additive noise.

In the case of telescope images, $A$ is commonly represented by a convolution with  a known PSF constrained from the images of stars, and a regular sampling by the detector. The  problem of image restoration is to recover an unknown source image $u_s$ from a given observed, blurred image $u_{0}$ and PSF $h$.  Eq.~\ref{Equ0_1}  takes the form:
 
\begin{equation}
u_{0} = h \otimes u_s + n,
\label{Equ0_2}
\end{equation}
where    $\otimes$ denotes the convolution operation, and h is the convolution  kernel representing the PSF of the telescope.

The deconvolution problem described above is a well-known ill-posed problem, which does not have a unique solution, as the convolution and the noise effectively destroy some information \citep{Alison1979}. One possible approach  for solving this problem is regularisation as the convolution and the noise effectively destroy the telescope images. A widely used regularisation method consists in minimising the Tikhonov functional \citep{Tikhonov1977}: 

\begin{equation}
{ J(u)} =  \|    u_{0} - h \otimes u \| ^2    +    \lambda  \, \|  D ( u ) \| ^2 \,
\label{Equ0_2a}
\end{equation}
where $\Arrowvert.\Arrowvert$  denotes the $L_2$ norm, and D is a differential operator.  The regularisation parameter  $\lambda>0$ is introduced to control the trade-off between the image bias and image variance \citep{Haykin2008}. The image that minimizes  the functional $J(u)$ above  is  an estimation of the intrinsic image $u_s$  (Eq.~\ref{Equ0_1}).

\subsection{Toeplitz matrix formulation}
\label{subsection:Toeplitz matrix}

In linear algebra, the convolution operator can be represented as a matrix-vector multiplication  by constructing a special Toeplitz matrix for a particular PSF kernel \citep{Toeplitz_2006}. The Tikhonov functional  $J(u)$ is expressed in a vector form as:

\begin{equation}
J({\mathbf u}) =
 \frac{1}{2} \|   {\mathbf  u_{0}} - H \, \bold{u} \| ^2    +   \frac{1}{2} \lambda  \, \|  D {\mathbf u }\| ^2,
\label{EqJu1}
\end{equation}

where $\bold{u}_{0}$ is the observed image and,  the variable $\bold{u}$  denotes  an approximation of the sours image $\bold{u}_{s}$, both represented as \emph{column vectors}, and $H$ is a Toeplitz matrix encoding the convolution. 

The convolution operation is now represented as a matrix multiplication, where known PSF image, $h(x,y),$ is converted into the corresponding Toeplitz matrix $H$.  A Toeplitz matrix is a sparse diagonal-constant matrix. If one considers the discrete convolution of a $n \times n$ image $I$ with a $n \times n$ PSF kernel $h$, the convolved image may be presented as

\begin{equation}
I_{conv}(x,y) = \sum_{x'=0}^{n-1}\sum_{y'=0}^{n-1} {I(x-x',y-y')}\,{h(x-x',y-y')}
\label{EqIconvxy1}
\end{equation} 
where $x$ and $y$ are the discrete indices of rows and columns, respectively, and $I_{conv}(x,y)$ is the result of the convolution operation for the pixel at position $(x,y)$.

\begin{figure}[t!]
\begin{center}
\includegraphics[scale=0.6  ]{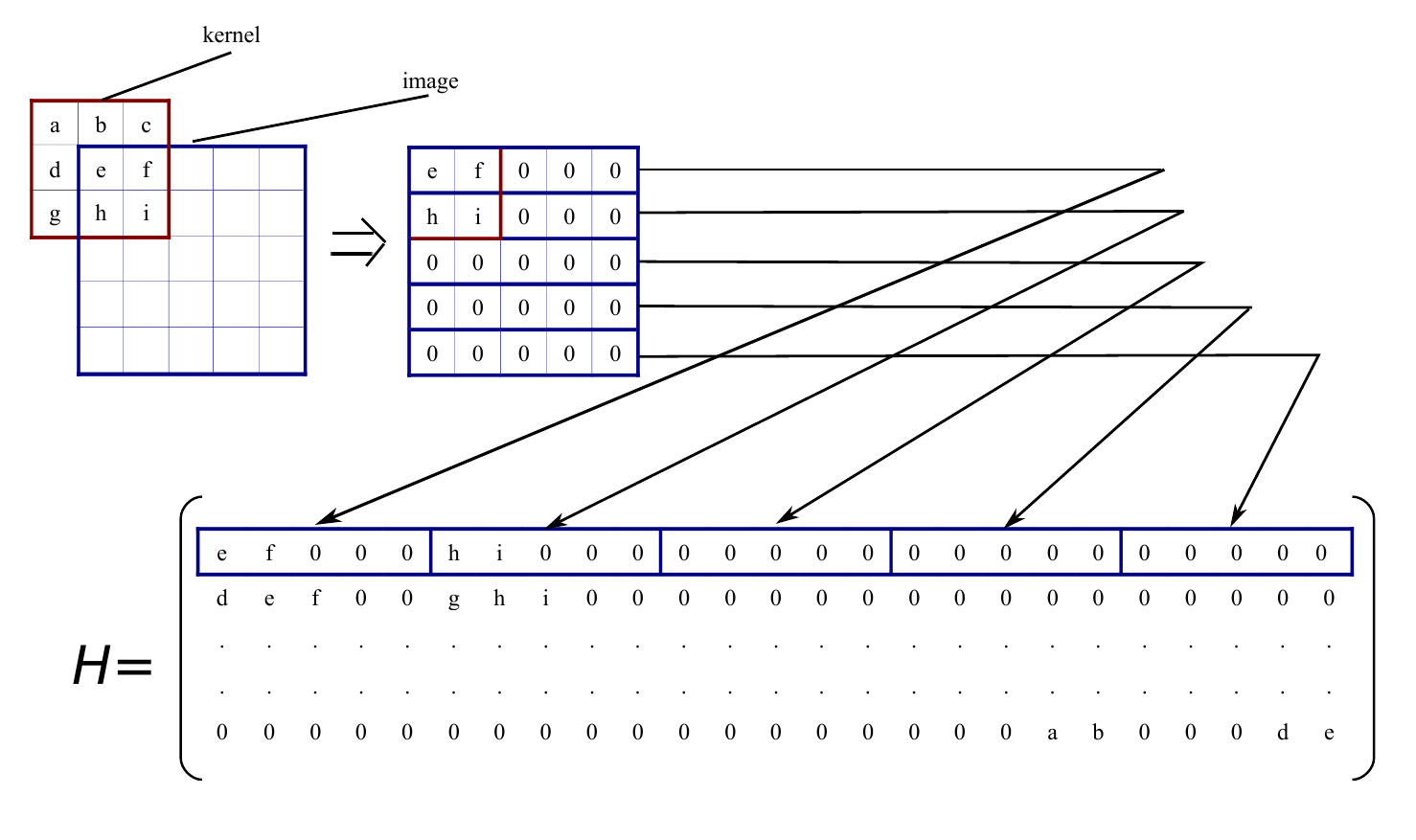}
\caption{Toeplitz matrix creation schema. A $5\times 5$ image  is convolvolved by a $3\times 3$ PSF kernel. The resulting Toeplitz matrix is $25$ pixels on a side}
\label{fig:Toeplitz}
\end{center}
\end{figure}

 Fig.~\ref{fig:Toeplitz} illustrates how to generate the first row of a Toeplitz matrix $H$ from a kernel $h$.  Each row of the Toeplitz matrix is a vector composed by ``flattened", row by row, kernel matrix coefficients corresponding to a particular pixel position in the image. For a $n\times n$ image, the corresponding Toeplitz matrix has a $n^2\times n^2$ dimension.

\subsection{Application of HNN to the image deconvolution problem}
\label{subsection:HNN}
\begin{figure*}[t!]
\begin{center}
\includegraphics[scale=0.6]{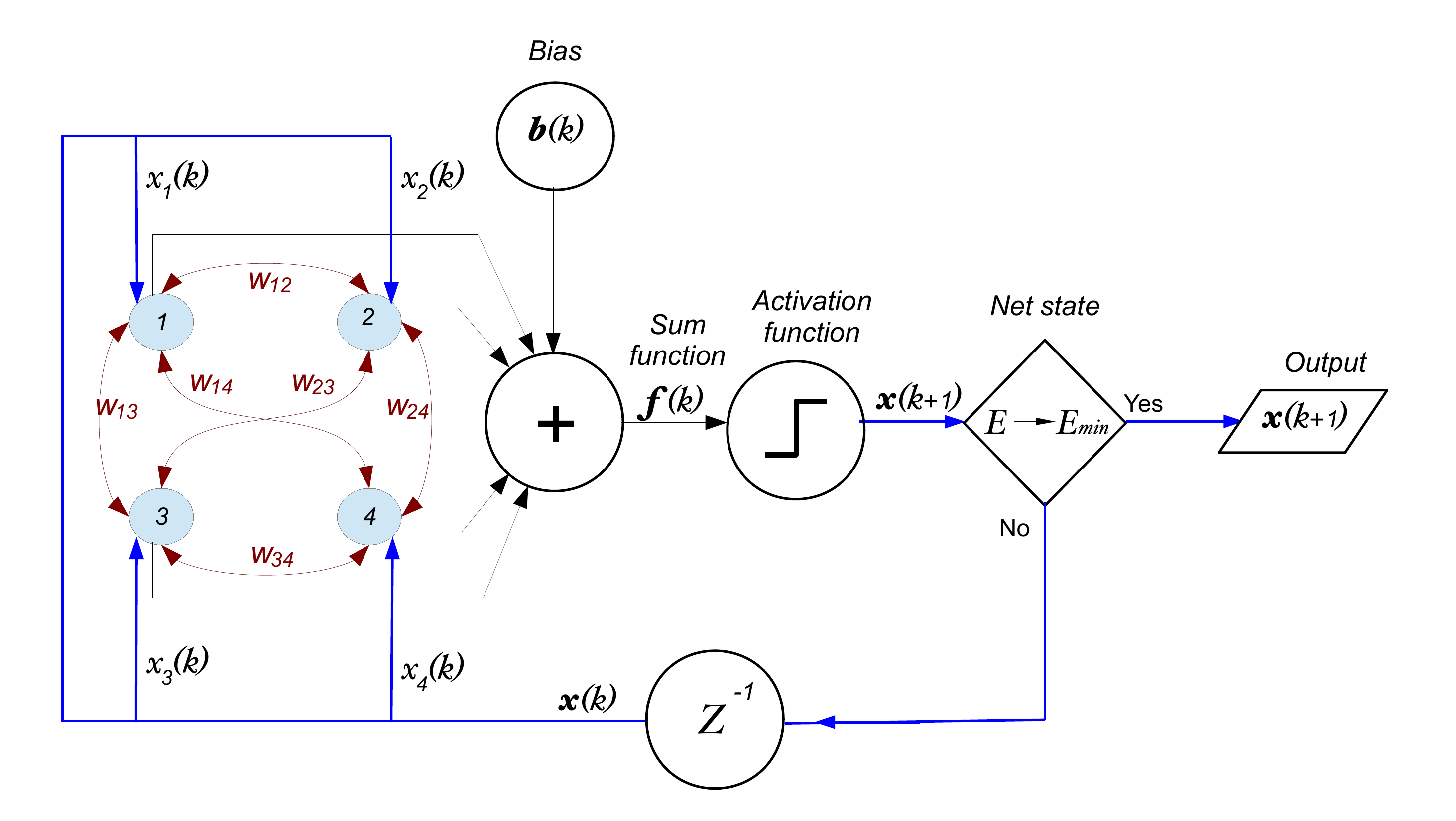}
\caption{Architecture of the Hopfield network of four neurons. The neurons are presented as light blue disks numbered 1, 2,3 and 4, $x_{i}(k)$, $(i = 1, 2, 3, 4)$ present the corresponding neuron state at time $k$ (iteration number). $W_{ij}$, $(i ,j = 1, 2, 3, 4)$ are the weights of interconnections between neurons. $b(k)$ denotes the bias at current time $k$. The output of the sum function $f(k)$  in iteration $k$ is the input vector for the activation function. The activation function uses the thresholding rule to calculate the neuron updates to change their states.  $Z^{-1}$ denotes the time delay of the iterative circuit. Since the network energy $E$ converges, the loop stops withe output vector  $y_{i}(k)$, $(i = 1, 2, 3, 4)$ }
\label{fig:HNN}
\end{center}
\end{figure*}

 The Hopfield neural network (HNN) model  is a widely-used approach in many different optimization problems  \citep{Hopfield1982,  Rojas1996}. Neural networks  are decision making tools operating on statistical data, inspired by the functioning of a biological nervous system. Each neural network model has a set of {\it neurons}   (nodes) characterised by the basic elements:
\begin{itemize}
\item {\it weights} (strength of the connections between neurons),
\item {\it sum function}  (calculates the total signal  incoming from all connected neurons),
\item {\it activation function} (calculates the net output).
\end{itemize} 

The principle of a HNN is shown on Fig.~\ref{fig:HNN} . The Hopfield model is a recurrent model, which consists of only one layer of neurons  all connected with each other, so that each neuron is the input and output at the same time, creating a self-feedback loop  with delayed connection from a neuron to itself. The weight matrix between neurons are symmetrical, e.g. $w_{ij}=w_{ji}$.  One of the advantages of HNN is that no prior information is necessary, so the neurons can be initialised with any non-negative values. 

The image reconstruction problem described above (Eq.~\ref{Equ0_2})  can be solved using a HNN whose neurons correspond to the pixels of the reconstructed image, as will be described in the following.  \citet{Sun2000} proposed the modified HNN model, where the activation function is a threshold function, determined by the General Updating Rule (GUR), which is theoretically proven to converge to the unique solution of Eq.~\ref{Equ0_2}.  According to the GUR, the updated value of the neuron  $i$ at  iteration {\it k}  is calculated as:

\begin{equation}
v_i(k)=
\left\{ 
\begin{array}{lll} 
v_i(k-1)- \Delta v_i(k), \qquad & f(k-1)<-\mathit{{TH}_i}\\
v_i(k-1)+ \Delta v_i(k), \qquad & f(k-1)>\,\mathit{{TH}_i}\\
v_i(k-1), \qquad & otherwise
\end{array}
\right. 
\label{update_rule1}
\end{equation}
where  $\Delta v_i(k)$ is the update value which equal the unit of the image brightness, and $TH$ and  $f$ are the threshold and sum function output, respectively. 

Hopfield nets have a scalar estimator associated with each iteration of the network referred to as the {\it energy function}, $E$ of the network. At each iteration the neurons are updated according to the GUR, decreasing the norm of the energy function.  The self-feedback loop of HNN stops at the stable state for the network - when there are no neurons to update according the activation function Eq.~\ref{update_rule1}. 

The energy function of HNN in a matrix form is defined as follows (\citet{Sun2000}):
\begin{equation}
E\bold{(u)} = \frac{1}{2} \, \bold{u}^T(W + \lambda \, G) \,\bold{u} - Q^T\bold{u}\ ,
\label{EqEnergy2}
\end{equation}
with

\begin{equation}
{W=H^TH}, \ G=D^TD,  \ Q=H^T\bold{u_{0}}\ .
\label{EqWGQ}
\end{equation} 
where $W$ is the synaptic weight matrix related to the PSF,  the  $Q$ term is associated with a bias vector of HNN and the combination $ (W + \lambda \, G)$ is the interconnection matrix between neurons, represented by image pixels. 

On the other hand, expanding  Eq.~\ref{EqJu1}, we have:
\begin{equation}
 J\bold{(u)} =
 \frac{1}{2}  {\bf u}^T (H^T H + \lambda  \,   D^TD){\bf u} - (H^T{\bf  u}_{0})^T {\bf u} +  \frac{1}{2}    {\bf u}_{0}^T  {\bf u}_{0},
\label{EqJu2}
\end{equation}
using  (\ref{EqWGQ}) in Eq. \ref{EqJu2}, we obtain

\begin{equation}
 J\bold{(u)} =
 \underbrace{ \frac{1}{2}  {\bf u}^T  (W + \lambda  \,   G){\bf u} - Q^T  {\bf u}  }_{\displaystyle E\bold{(u)}} +  \frac{1}{2}    {\bf u}_{0}^T  {\bf u}_{0} 
\label{EqJu3}
\end{equation}

One can see that   the Tikhonov regularization functional $J\bold{(u)}$  and the energy function of the HNN network $E\bold{(u)}$ differ only  in the term $\frac{1}{2}  {\bf u}_0^T {\bf u}_{0} $ which doesn't depend on ${\bf u}$. Therefore, by minimizing the HNN energy  we can find the best estimation of a solution of  Eq.~\ref{Equ0_1}  

The network is run as a series of iterative steps where the values of selected neurons (pixels) are updated so as to minimize the energy of the whole network. Image pixels are updated according to the condition Eq. \ref{update_rule1}, where the threshold value is given by

\begin{equation}
\mathit{{TH}_i} =  \frac{1}{2}  \,  \big\arrowvert W_{ii}+  \lambda \, G_{ii} \big\arrowvert  \, , \\ i=1, .., N^2
\label{EqTH}
\end{equation}
At each iteration the pixels exceeding the threshold are updated, minimizing the energy function. The network stops running when its total energy $E$ reaches a global minimum, that is, when no pixel needs to be updated the image is assumed to be fully deconvolved.

\subsection{ Modification of the Hopfield Neural Network: adaptive neuron updating rule }
\label{subsection:neuron update}

In the application of HNN to natural images the updating value $\Delta v_i(k)$ is an integer constant, usually equal to 1 (for instance, for the images where the pixel value varies from 1 to 256). However, astronomical images have a great range of pixel intensities. Such an image can be divided into two regions, the background area that contains very little signal, and the regions of astronomical objects themselves, where the dynamical range  is  large. These special properties of astronomical images must be taken into account to determine an optimal updating value.

For the HNN deconvolution the selection of an optimal pixel updating step is very important as:

\begin{itemize}
\item a small updating value requires a large number of iterations, making the minimisation process very slow;
\item a large updating value can result in an instability of the HNN energy function, preventing the neural network from converging.
\end{itemize}

We propose a new adaptive neuron updating rule, which takes into account the specific properties of  astronomical images and that accelerates the HNN deconvolution process. At each iteration $k$, the neuron update  for a given pixel   $ \Delta v_i(k)$,  is proportional to the gradient of the energy function, which allows to reduce the difference between the observed image and the image being reconstructed convolved with the PSF kernel:

\begin{equation}
\Delta v_i(k) =  c \big\arrowvert \nabla{E(v(k))} \big\arrowvert   =  c\, [ (W+\lambda \, G) \,v(k)-Q],
\label{delta_x}
\end{equation} 
where $c$ is a constant which scales the updating values, and therefore also influences the number of required iterations.. The set of neuron updating values $\Delta v(k)$ is an array, where each element is associated with certain pixel in the image.

\section{ Galaxy shape measurement }
\label{section:Galaxy shape measurement}

The galaxy shape measurement process consists of three main steps:

\begin{enumerate}
	\item  Galaxy image reconstruction by iterative application of the HNN image deconvolution algorithm (Sect.~\ref{subsection:deconv_scheme}) to ``postage stamps" containing a single galaxy image each. 
	\item Computation of the two-dimensional ACF for each deconvolved galaxy image (Sect.~\ref{subsection:ACF}). 
	\item Measurement of the galaxy ellipticities by calculating the quadrupole moments of their ACF images (Sect.~\ref{subsection:ellipticities}).
\end{enumerate}

\subsection{Image deconvolution scheme}
\label{subsection:deconv_scheme}

The deconvolution algorithm presented in Sect.~\ref{section:method} is implemented in the following way:

{We represent each galaxy postage stamp $I$ of dimension $n \times n$ pixels, as a ``flattened" vector $v$ (Sect.~\ref{section:method}).  For each galaxy vector $v$ we:}

\begin{itemize}
\item{Generate a vector $y$ of length $n^2$ by stacking the rows of the observed image  $u_0$;}
\item{Generate the Toeplitz matrix $H$ using the PSF kernel \citep{Toeplitz_2006} and the matrix $D$ using a high pass filter $d$, designed such that the summation of $d_{ij}$ is equal to zero. In our method we use the kernel $d$ proposed by \citet{Sun2000}  :

\begin{displaymath}
d= \frac{1}{100}  \,\left( \begin{array}{ccc}
0.7 & 1 & 0.7 \\
1 & -6.8 & 1 \\
0.7 & 1 & 0.7 \\
\end{array} \right);
\end{displaymath}
}

\item{Calculate the matrices $W$,  $G$ and  $Q$ (Eq.~\ref{EqWGQ}),}

\item{Set $c=0.001*max(u_0)$ and $\lambda=0.0001 ()$}

\item{Set $k=0$  \space and \space $X(0)=Y$.}

\item{Compute the threshold vector according to Eq.~\ref{EqTH}}

\item{Set $k=1$ and repeat the following steps while $k<M$:}

\begin{enumerate}

\item{ Calculate the negative gradient vector of the energy function (Eq.~\ref{delta_x}): }

\begin{equation}
f(k) =  Q - (W+\lambda \, G) \,v(k)
\label{NN_f}
\end{equation} 

\item{Compute the neuron updating vector $\Delta v(k)$  according to Eq.~\ref{delta_x}} 

\item Update the vector $v(k)$ according to the GUR  (Eq.~\ref{update_rule1}). If HNN stopped at the stable state, e.g. there is no pixel to update, then break the iteration loop. 

\item Correct the vector $v(k)$ as follows:
\begin{equation}
v_i(k)=
\left\{ 
\begin{array}{lll} 
0,\qquad & v_i(k)<0\\
v_i(k),\qquad & otherwise
\end{array}
\right. 
\label{update_rule}
\end{equation}

\item{Check that $E(k+1)<E(k)$, otherwise break the iteration loop.}
\end{enumerate}
\item{Reshape the flattened vector $v(k)$ into a two-dimensional image $I^\star$. }

\end{itemize}

\subsection{The autocorrelation function in a shape measurement}
\label{subsection:ACF}

\begin{figure}
\begin{center}
\leavevmode
\includegraphics[scale=0.17]{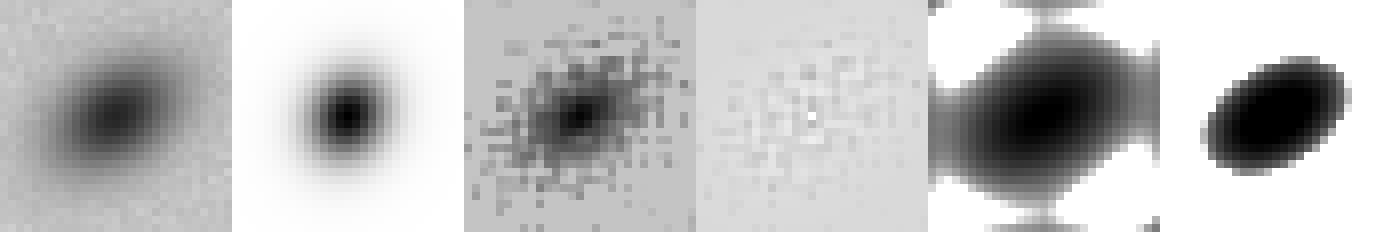}
\includegraphics[scale=0.17]{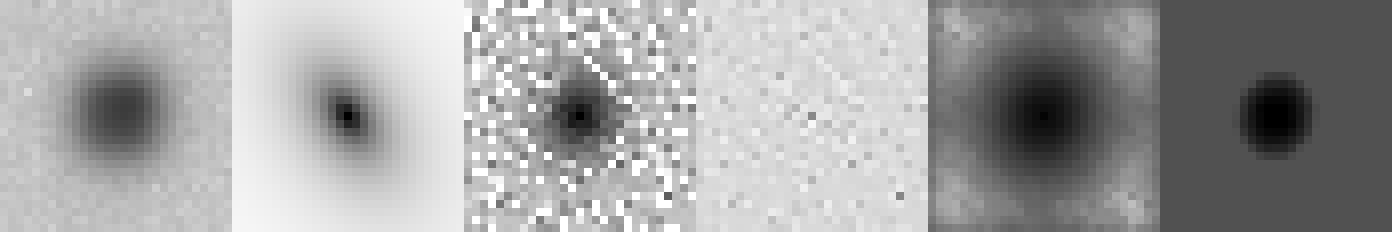}
\includegraphics[scale=0.17]{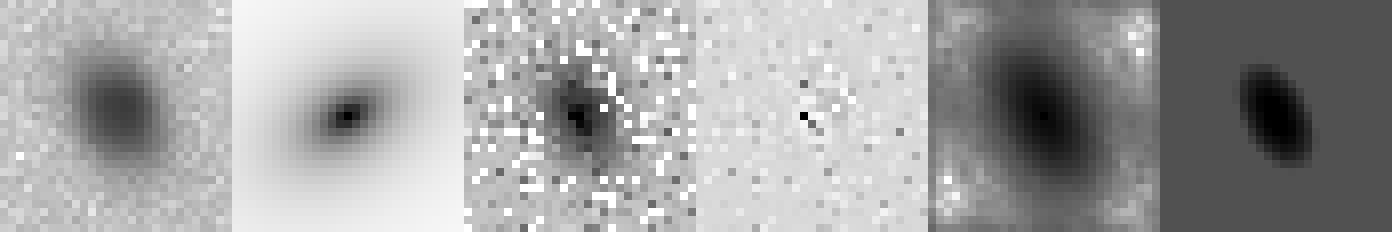}
\caption{Illustration of the HNN deconvolution, using galaxies from the GREAT10 image processing challenge. The three rows show three different galaxies. On each row, are displayed from left to right: the original GREAT10 postage stamp for each galaxy, the associated PSF, the HNN deconvolution, the residual image (difference between data and deconvolved images, reconvolved by the PSF), the ACF of the deconvolved galaxy, and the ACF clipped at the radius that encloses 90\% of the total light.}
\label{fig:GalDeconvolution}
\end{center}
\end{figure}

Once the deconvolved galaxies, $I^\star$, have been obtained, the next step is to measure their ellipticities. Our goal with the HNN deconvolution method is to allow shape measurements free of any underlying model for the light distribution of galaxies. Therefore the ellipticity measurement has  to be carried out in a non-parametric way, e.g., using the 2nd order moments of the light distribution of $I^\star$. 

In practice, such a measurement is sensitive to noise and requires the estimation of the centroid of the galaxies. A safe and convenient way to circumvent both limitations is to measure the galaxy ellipticities using the ACF of the light distribution and not the galaxy images directly. Indeed, it has been shown by \citet{Miralda-Escude1991} and proven later by \citet{Waerbeke1997} that the ellipticity of the ACF is the same as the ellipticity of the image itself. In addition, by construction, the ACF of any function is exactly centered at the center of the frame. We therefore measure the ellipticity of the galaxy ACF rather than the ellipticity of the galaxy itself. 

Fig.~\ref{fig:GalDeconvolution} shows examples of galaxies borrowed from the GREAT10 image processing challenge, along with their HNN deconvolution and ACF functions. To minimize edge effects caused by the square nature of the galaxy stamps, the ACF is  clipped at the radius that encloses 90\% of the total light. Such a cutoff radius removes nearly all edge effects and improves the calculation of the ellipticity through the second-order moments method.
 
\subsection{Estimation of galaxy ellipticities}
\label{subsection:ellipticities}

We use the ACF of the deconvolved galaxy images, $I^\star$, to measure their complex ellipticity, $e = e_1 + i\, e_2$. The ellipticity components ($e_1, e_2$) are derived from the unweighted quadrupole moments $Q$, calculated around the center of the light distribution ($x_c, y_c$):

\begin{equation}
\label{ellipticity from quadrupoles}
e_1 = \frac{Q_{xx} - Q_{yy}}{Q_{xx} + Q_{yy}}, \qquad e_2 = \frac{2\, Q_{xy}}{Q_{xx} + Q_{yy}},
\end{equation} 
where

\begin{equation}\label{quadrupole moments}
 \begin{array}{l}
  Q_{xx} = \int{I^*(x,y)\,\, (x-x_c)^2 \, dx \,\, dy}, \\
  Q_{xy} = \int{I^*(x,y)\,\, (x-x_c) \, (y-y_c)\,\, dx \, dy}, \\
  Q_{yy} = \int{I^*(x,y)\,\, (y-y_c)^2 \,\, dx \, dy}. \\
 \end{array} 
\end{equation} 
In the present case, we use the ACF of the light distribution, so ($x_c, y_c$) are known exactly and coincide with the center of the image.

\section{ Tikhonov regularization parameter}
\label{section:Parameter_lambda}

\hspace{3ex}  The important question in Tikhonov regularization is how to determine the optimal parameter  $\lambda > 0$ in order to find a solution $u^*$ close to the true noise-free solution $u$. The parameter $\lambda$  regulates the balance between the accuracy decreasing with $\lambda$, and the smoothness of the restored image increasing with $\lambda$. In order to find an optimal value of $\lambda$   we applied the  image deconvolution method described above to the synthetic data.  Hereafter we discuss the calibration of $\lambda$ based on simulations.

Each synthetic  dataset consists of 10000  galaxy images of a S\'ersic profile  convolved  with asimulated Moffat-profile PSF. The  noise added to the blurred images has a Gaussian distribution with mean zero and the
corresponding signal-to-noise ratio (SNR) ranges from 10 to 40 dB. The SNR of the convolved noisy image is estimated as 
$S\!NR=10 \ log_{10}\ (\sigma_{S}^2 / \sigma^2)$, where $\sigma_{S}^2 $ is a variance from the noise-free convolved image and $\sigma^2$ is the variance of the noise added to the convolved image. 

\begin{figure}
\begin{center}
\includegraphics[scale=0.44]{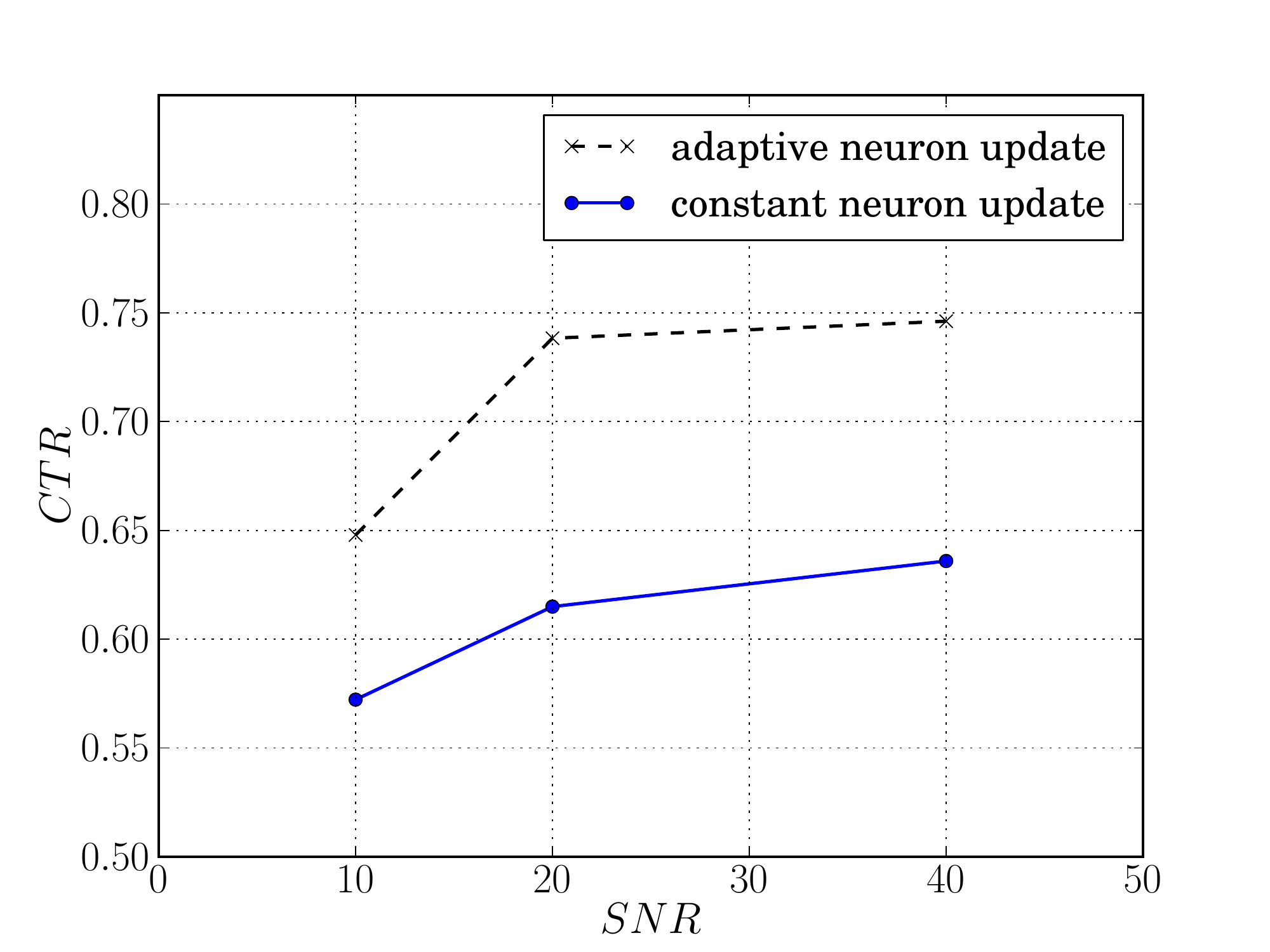}
\caption{Correct transition rate as a function of  signal-to-noise ratio for the adaptive update approach (black dashed line) and contents update approach (blue solid line).  $\lambda=10^{-3.5}$ }
\label{fig:CTRvrSNR}
\end{center}
\end{figure}

To evaluate the efficiency of our algorithm we use the metrics called Correct Transition Rate (CTR) and Discriminative Signal-to-Noise Ratio (dSNR) described below  (\citet{Sun2000, Ben-Arie1993}).
To recover the individual galaxy image a certain number of iterations is required. In each iteration, the states of some neurons change according the GUR. If the state transition for a given neuron $j$ decreases the difference between the state of the neuron and the corresponding pixel value of  the original image $\arrowvert  v^{j}(k) - u^{j} \arrowvert$, the transition is called {\it correct transition}. The CTR is defined as:

\begin{equation}
\label{CTR}
CTR = \frac{K_{c}}{K_{t}}.
\end{equation} 
where $K_{c}$ is the number of the correct transitions, and $K_{t}$ is total number of the neural  state transitions. The CTR varies from 0 to 1 and measures the convergence rate of the algorithm. The higher the CTR is, the better the efficiency of the algorithm. 

Fig.~\ref{fig:CTRvrSNR} shows the CTR factor for both the proposed adaptive neuron updating and standard constant neuron updating. One can see that, whatever the SNR, the  adaptive neuron updating approach demonstrates a higher convergence rate compared to the  constant updating.  

\begin{figure}
\begin{center}
\includegraphics[scale=0.44]{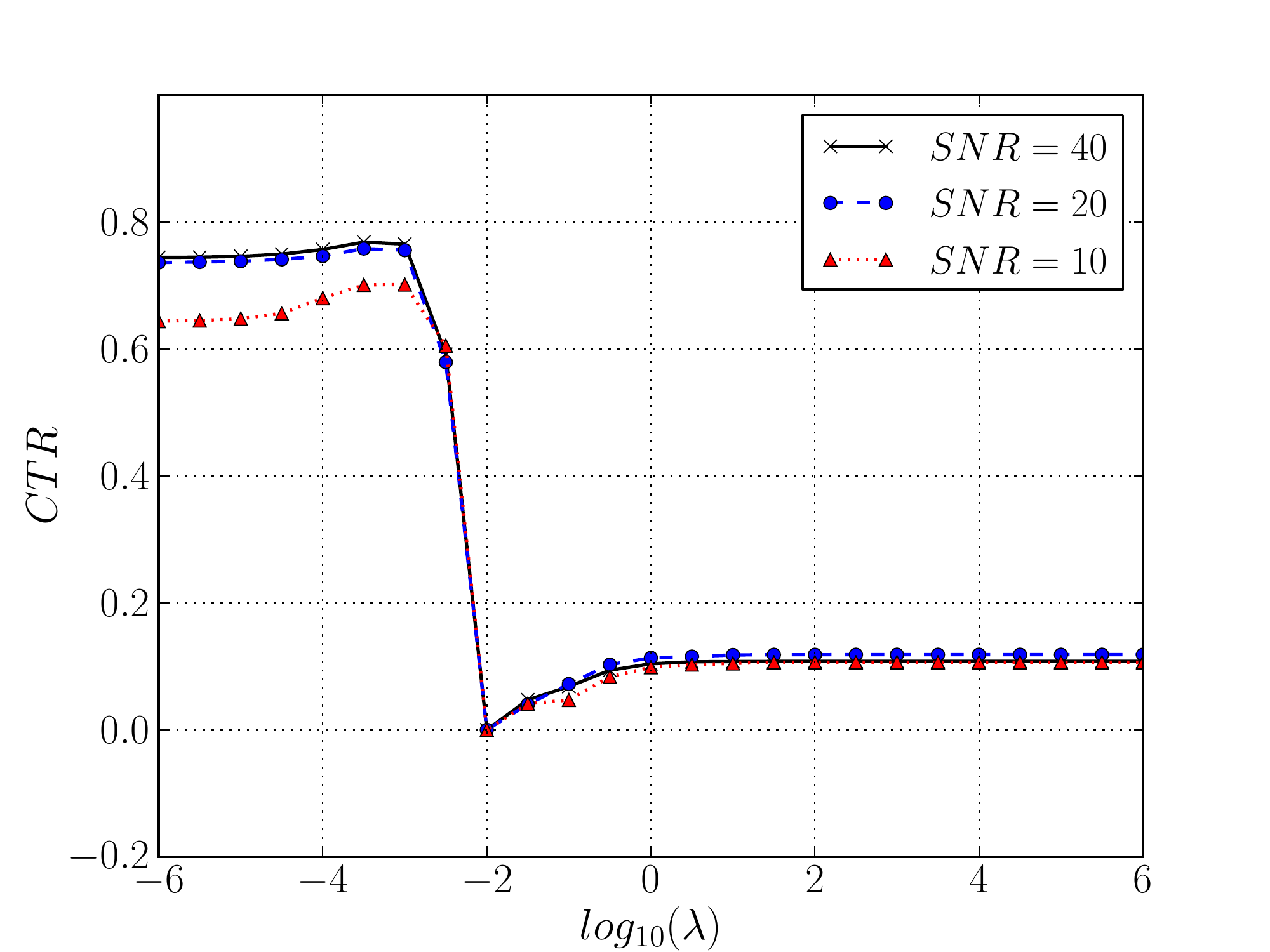}
\caption{Correct transition rate as a function of the  Tikhonov regularisation  parameter $\lambda$ for $S\!NR=10$ (red dotted line),  $S\!NR=20$ (blue dashed line), $S\!NR=40$ (black solid line).}
\label{fig:CTRvrLambda}
\end{center}
\end{figure}

\begin{figure}
\begin{center}
\includegraphics[scale=0.44]{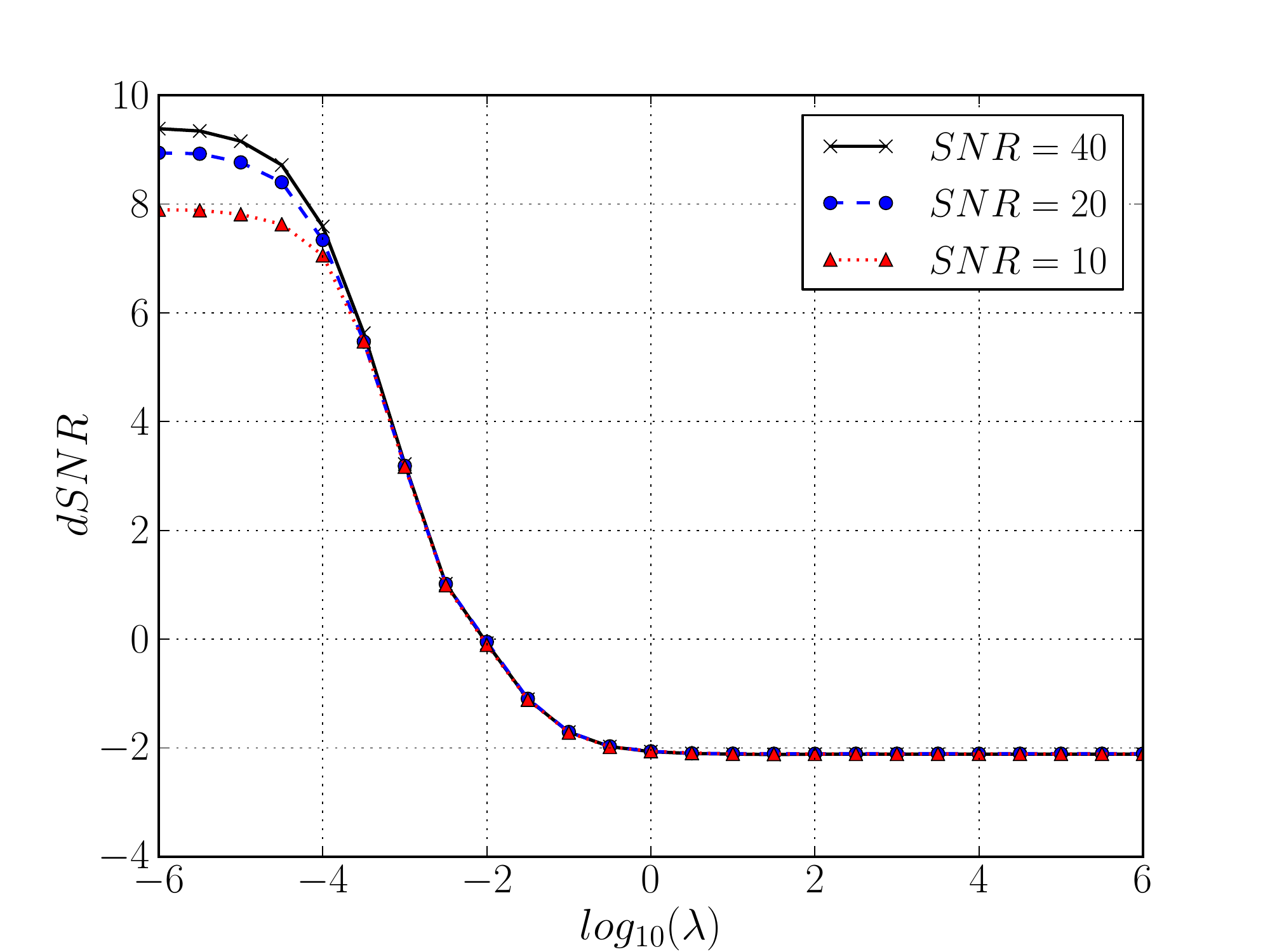}
\caption{Discriptive signal-to-ratio as a function of the   Tikhonov regularisation parameter $\lambda$  for $S\!NR=10$ (red dotted line),  $S\!NR=20$ (blue dashed line), $S\!NR=40$ (black solid line).}
\label{fig:dSNRmodified.png}
\end{center}
\end{figure}

\begin{figure}

\centering
\includegraphics[width=0.9\linewidth]{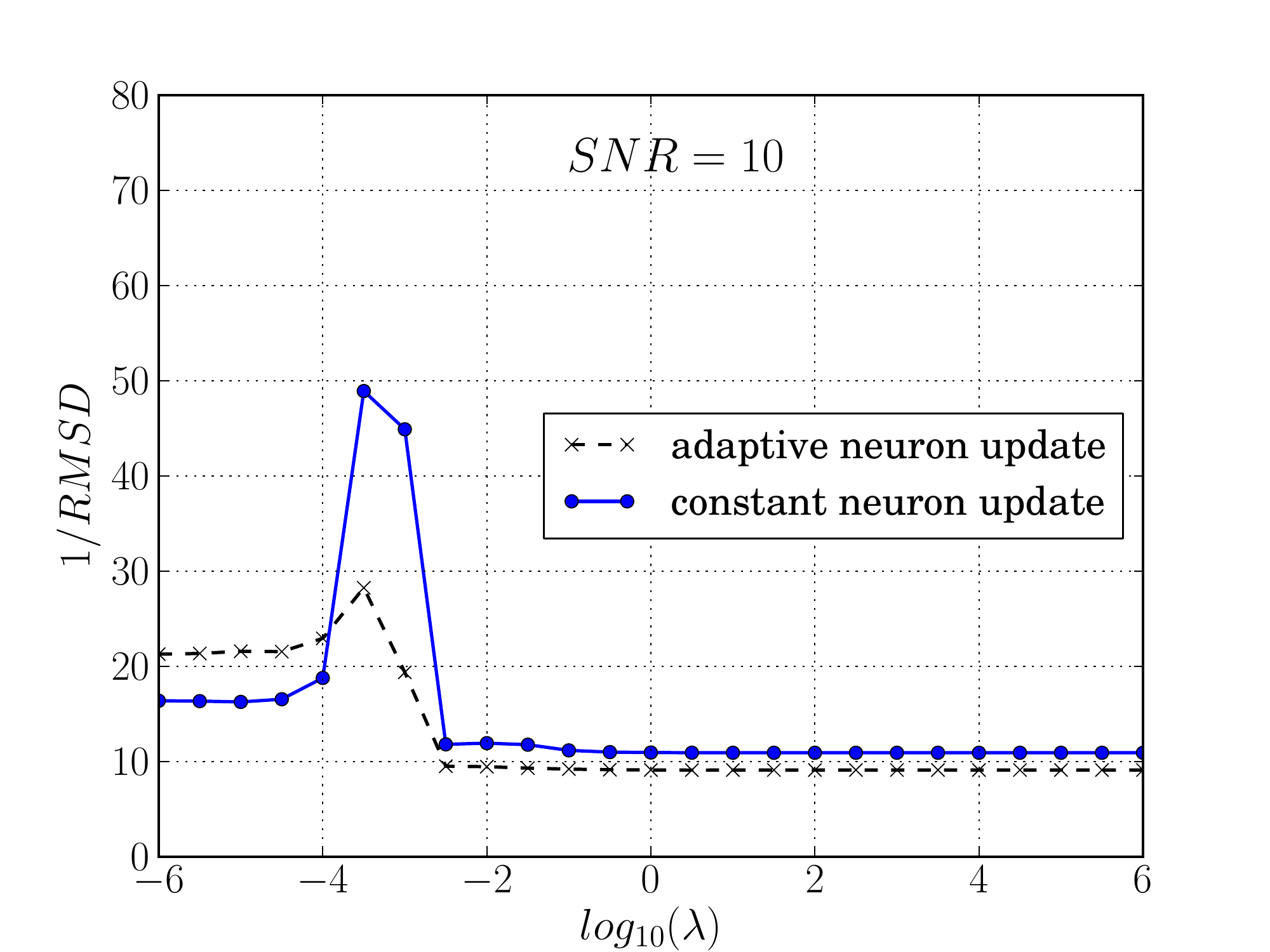}
\includegraphics[width=0.9\linewidth]{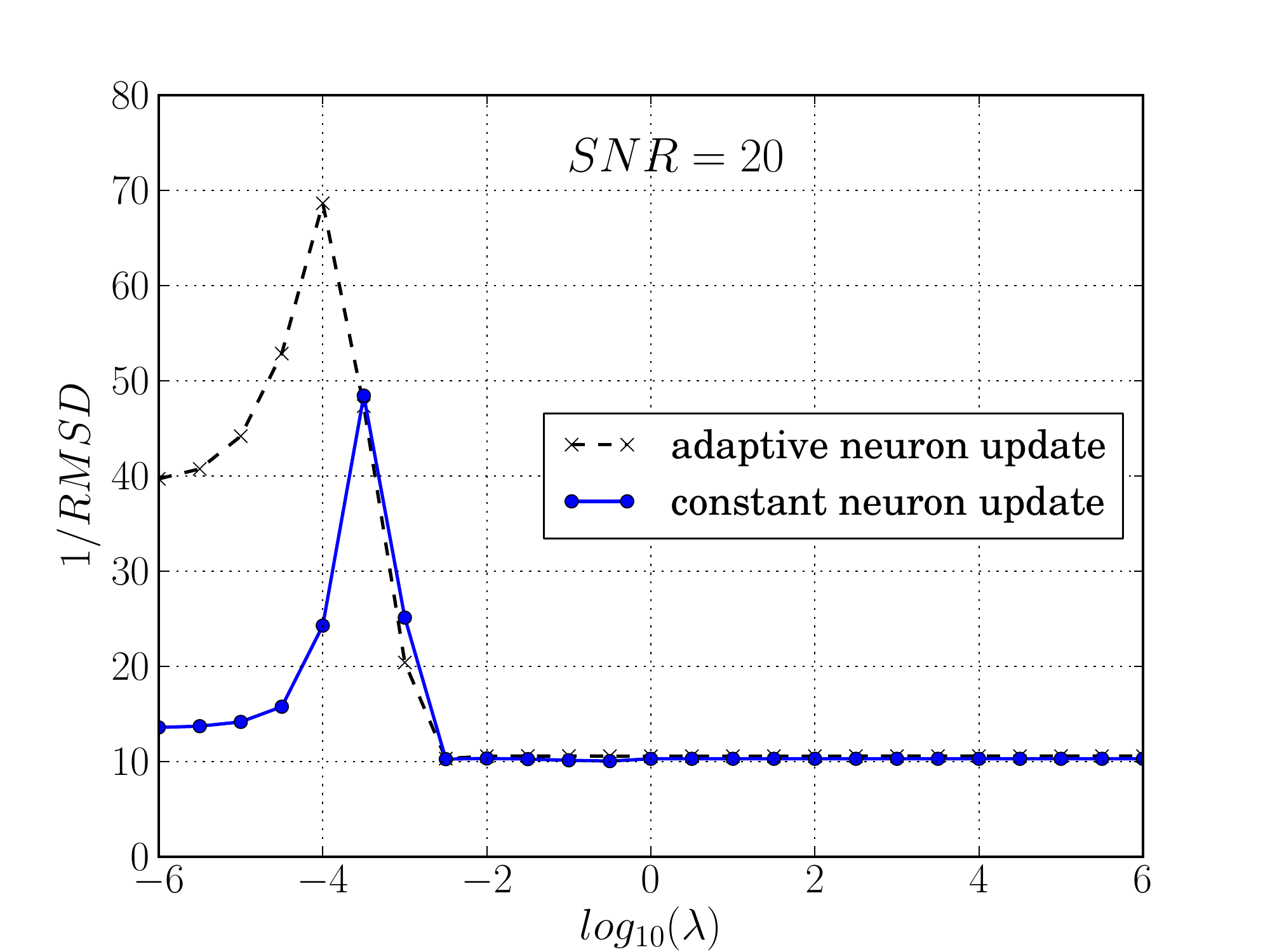}
\includegraphics[width=0.9\linewidth]{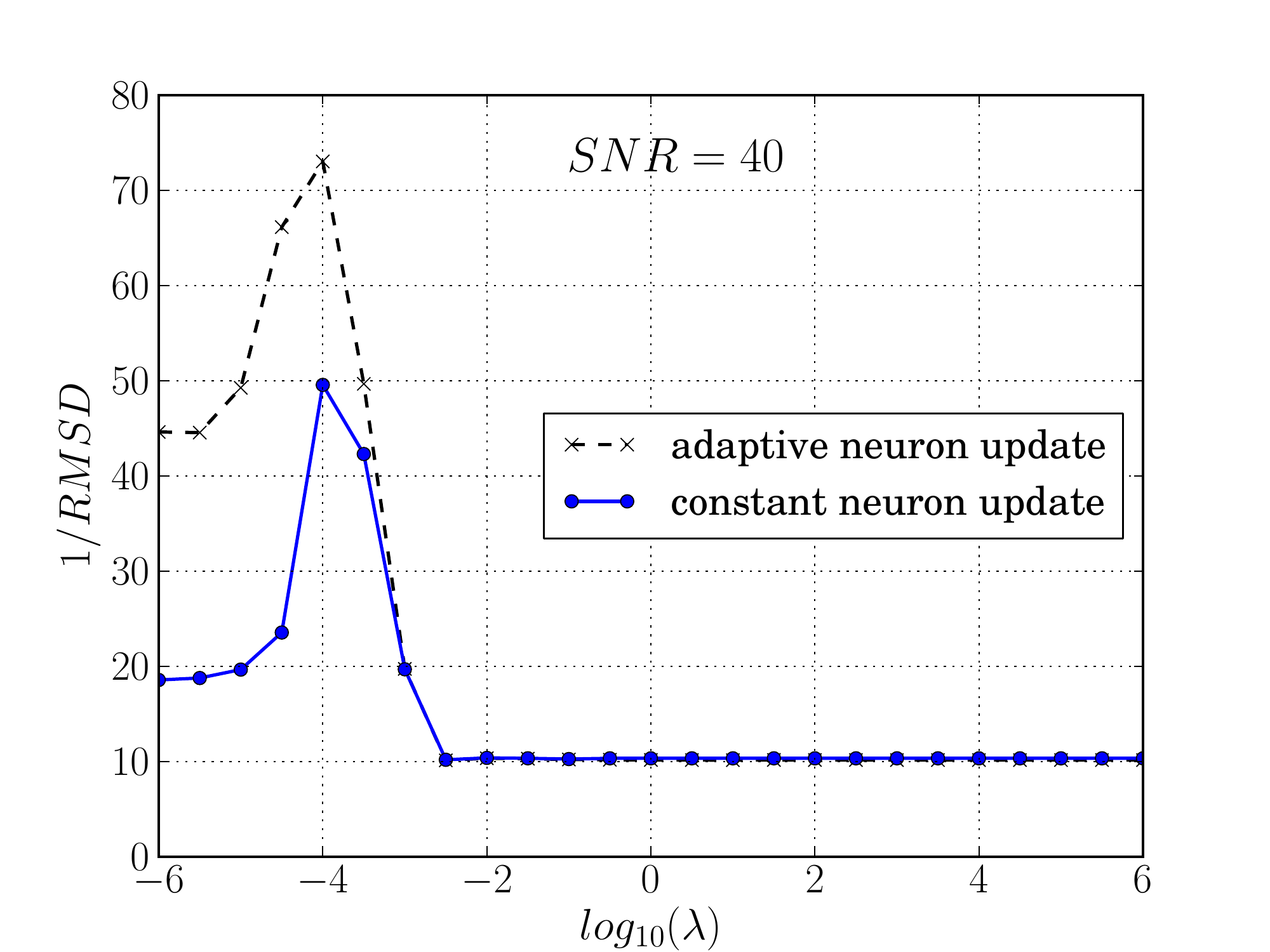}
\caption{Accuracy on the shape measurement (as inverse RMSD) as a function of the Tikhonov regularisation parameter $\lambda$ for two different methods: the  adaptive neuron update  (black dashed line) and constant neuron update (blue solid line). Figures from top to the bottom correspond to SNR equal 10, 20 and 40 dB, respectively.  }
\label{fig:RMSD_SNR}
\end{figure}

The CTR values for  $\lambda$  varying from $10^{-6}$ to $10^{6}$ is  shown in Fig.~\ref{fig:CTRvrLambda} the CTR increases with SNR, though for ${\lambda>10^{-2.5}}$ the difference on CTR is negligible. The highest CTR value corresponds   to  $\lambda=10^{-3.5}$. 

To estimate the performance of the image reconstruction, we use the dSNR criterion estimated as follows:
\begin{equation}
 dS\!NR   \, \equiv 10 \, log_{10}  \left( \frac{\|     u_{0} - u \| ^2 } {  \|  u^* - u \| ^2 } \right),
\label{dSNR}
\end{equation}
where $u$ is the original simulated galaxy image, $u^*$ is the final restored image and $u_0$ denotes the convolved noisy image.  As shown in Fig.~\ref{fig:dSNRmodified.png},   dSNR is higher for  higher $S\!NR$ and decreases with $\lambda$.

The key parameter which controls the performance of the proposed approach is the accuracy on shape measurement. We measured the ellipticities  of the reconstructed galaxies using the technique explained in Sect.~\ref{section:Galaxy shape measurement}. The accuracy on shape measurement  is estimated using the inverse Root Mean Square Deviation (RMSD):
\begin{equation}
1/RMS\!D   \, \equiv  \sqrt { \frac{2N } {    \sum_{i=1}^{N} {(e_{1,i}^{ } - e_{1,i}^*)}^2 + {(e_{2,i}^{ } -e_{2,i}^*)^2}   } }
\label{RMSD}
\end{equation}
where $e_{1,i}^{ }$, $e_{2,i}^{ }$ are the intrinsic ellipticities and  $e_{1,i}^*$, $e_{2,i}^*$ are the estimated ellipticities calculated from  (Eq.~\ref{ellipticity from quadrupoles}). $i$ denotes the galaxy number in the set with $N=10000$ galaxies.

Fig.~\ref{fig:RMSD_SNR} shows the accuracy on the shape measurement for both adaptive neuron update  and constant  neuron update.  One can see that for $\lambda  \leq 10^{-4}$  the adaptive neuron update gives better accuracy regardless the noise level of the images. For $\lambda  > 10^{-4}$ the accuracy is sensitive to the noise level of the images. While on SNR level of 20 and 40 the adaptive neuron update still shows  better performance, on noisier images with $S\!NR=10$ it works less well than with constant neuron updating.

Taking into account the CPU time and the quality of the reconstructed images (Fig.~\ref{fig:CTRvrSNR}, \ref{fig:CTRvrLambda}, \ref{fig:dSNRmodified.png}, \ref{fig:RMSD_SNR}) the optimal value of $\lambda=10^{-4}$ is suggested.

\section{Application to the GREAT10 data}
\label{section:application to the GREAT10 data}

\subsection{The GREAT10 galaxy challenge and its dataset}
\label{subsection:GREAT10 dataset}

The image deconvolution scheme described in the paper was used to participate in the Gravitational LEnsing Accuracy Testing 2010 challenge (GREAT10), an image processing competition that ran from December 2010 to September 2011  \citep{GREAT10Handbook2010, G10results} and whose goals are (i) to test existing shear measurements methods and (ii) to promote the development of new shear measurement techniques. The GREAT10 challenge is the continuation of the GREAT08 challenge \citep{BridleGREAT08Handbook2008, BridleGREAT082010} and STEP programs \citep{HeymansSTEPI2006, MasseySTEP22007, BridleGREAT082010}, with an increasing degree of complexity.

\begin{table*}
\center
\caption{Q factors obtained by the different versions of HNN deconvolution. The first four columns refer respectively to the method version used, the size of the postage stamps cutouts, whether the data were denoised (den) or not (raw), and whether a functional form was used for the PSF (func) or a star postage stamp (star). The $Q$ column denotes the quality factor as originally published in the leadeboard.  The  $\mathcal{M}/2$ and $\sqrt{\mathcal{A}}/10^{-4}$ columns indicate respectively the average multiplicative and additive biases over all image sets. The NN23 methods was submitted in the Post-challenge, after the official galaxy challenge deadline.}
\renewcommand{\arraystretch}{1.10}
\label{table:Q factor per method}
\begin{tabular}{lccrrrr}
\hline
Method & Size & Den & PSF &$\mathcal{M}/2$ & $\sqrt{\mathcal{A}}\times 10^{-4}$ & $Q$ \\
\hline \hline
Subm. 19/den/star\tablefootmark{\dagger}  & $19\times 19$ & Y & Star &  $0.0626$ & $-0.1500$ & $17.78$\\
Subm. 23/den/func\tablefootmark{\dagger} & $23\times 23$ & Y  & Func &  $-0.0153$ & $0.0982$ & 83.16 \\
23/den/func\tablefootmark{\dagger \dagger}  & $23\times 23$ & Y & Func & $-0.0364$ & $0.0900$ & 83.88\\
23/den/star\tablefootmark{\dagger \dagger}  & $23\times 23$ & Y & Star & $  0.0357  $ & $  0.0895 $ & 87.06 \\
23/raw/func\tablefootmark{\dagger \dagger} & $23\times 23$  & N & Func & $ 0.0218$  & $0.1030$ & 71.62\\
23/raw/star\tablefootmark{\dagger \dagger}  & $23\times 23$ &  N & Star & $-0.0239$  & $0.1040$ & 80.33\\
29/den/func\tablefootmark{\dagger \dagger} & $29\times 29$ & Y & Func & $0.0416$ & $0.0771$ & 90,70\\
\hline
\end{tabular}
\tablefoot{
\tablefoottext{\dagger}{Submitted in the Post-Challenge,} \
\tablefoottext{\dagger \dagger}{New results}
}
\end{table*}

\definecolor{FullVariability}{rgb}{0.824,0.824,0.824}
\begin{table}
\caption{ PSF and galaxy properties of the GREAT10 image sets. The second and third columns specify whether the PSF or intrinsic ellipticity field are the same for all 200 images within a set. The fourth column has the galaxy parameters ( \citet{G10results}). The default SNR is 20, while low and high SNR ratios are 10 and 40 respectively.  Sets 1-20 have galaxies with co-centered bulges and disks with a 50/50 bulge-to-disk ratio.
}
\renewcommand{\arraystretch}{1.2}
\label{table:property per set}

\begin{tabular}{llll}
\hline
Set & PSF & Ellipticity & Image type \\ 
\hline
\cellcolor{FullVariability}1 & \cellcolor{FullVariability}Variable &\cellcolor{FullVariability}Variable &\cellcolor{FullVariability}Fiducial  \\
2 & Fixed & Variable & Fiducial  \\
3 & Variable & Fixed & Fiducial \\
\cellcolor{FullVariability}4 & \cellcolor{FullVariability}Variable &\cellcolor{FullVariability}Variable &\cellcolor{FullVariability}Low S/N  \\
5 & Fixed & Variable &  Low S/N \\
6 & Variable & Fixed &  Low S/N  \\
\cellcolor{FullVariability}7 & \cellcolor{FullVariability}Variable &\cellcolor{FullVariability}Variable &\cellcolor{FullVariability}High S/N  \\
8 & Fixed & Variable &  High S/N  \\
9 & Variable & Fixed & High S/N  \\
\cellcolor{FullVariability}10 & \cellcolor{FullVariability}Variable &\cellcolor{FullVariability}Variable &\cellcolor{FullVariability}Smooth S/N  \\
11 & Fixed & Variable &  Smooth S/N  \\
12 & Variable & Fixed & Smooth S/N  \\
\cellcolor{FullVariability}13 & \cellcolor{FullVariability}Variable &\cellcolor{FullVariability}Variable &\cellcolor{FullVariability}Small galaxy  \\
14 & Fixed & Variable & Small galaxy \\
\cellcolor{FullVariability}15 & \cellcolor{FullVariability}Variable &\cellcolor{FullVariability}Variable &\cellcolor{FullVariability}Large galaxy  \\
16 & Fixed & Variable &  Large galaxy \\
\cellcolor{FullVariability}17 & \cellcolor{FullVariability}Variable &\cellcolor{FullVariability}Variable &\cellcolor{FullVariability}Smooth galaxy  \\
18 & Fixed & Variable &  Smooth galaxy \\
\cellcolor{FullVariability}19 & \cellcolor{FullVariability}Variable &\cellcolor{FullVariability}Variable &\cellcolor{FullVariability}Kolmogorov PSF  \\
20 & Fixed & Variable & Kolmogorov PSF \\
\cellcolor{FullVariability}21 & \cellcolor{FullVariability}Variable &\cellcolor{FullVariability}Variable &\cellcolor{FullVariability}Uniform bulge/disc ratios \\
22 & Fixed & Variable &  Uniform bulge/disc ratios \\
\cellcolor{FullVariability}23 & \cellcolor{FullVariability}Variable &\cellcolor{FullVariability}Variable &\cellcolor{FullVariability}50/50 bulge/disc offset\\
24 & Fixed & Variable & 50/50 bulge/disc offset \\
\hline
\end{tabular}
\end{table}

\begin{figure}[t!]
\begin{center}
\includegraphics[scale=0.68]{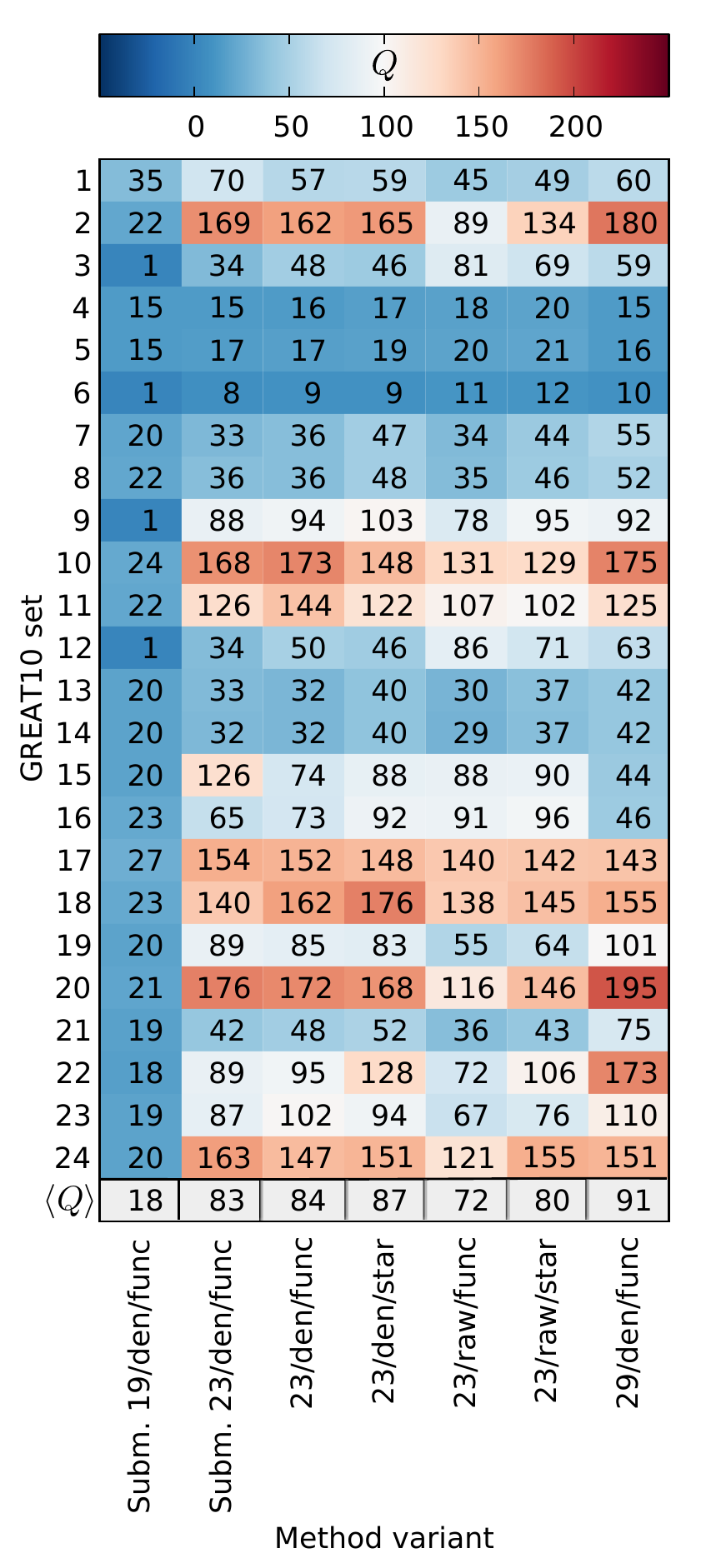}
\caption{Q factor comparison for all 24 sets of GREAT10 data for different method variants (see Table~\ref{table:Q factor per method}). The last line give the mean $Q$ factor for all 24 sets.}
\label{fig:fig_Qcomparision}
\end{center}
\end{figure}

The data of the GREAT10 challenge consist of 24 datasets of 200 simulated galaxy fields. Each image contains 10000 noisy, PSF-convolved galaxy images, arranged on a $100 \times 100$ grid. Each galaxy fliad is an image stamp of $48 \times 48$ pixels.

The main innovation in the 2010 edition of the challenge is the inclusion of variable components for the intrinsic ellipticities for the galaxies and of the shear fields. The PSF fields are also spatially varying and are provided to the challenge participants both as functional form and as FITS images. 

With the simulated data, participants to the challenge must measure the ellipticities of the galaxies, as they were prior to the convolution by the PSF and the addition of noise. The results are submitted to the challenge organizers either under the form of a catalogue of ellipticities or as a shear power spectrum. A ``quality factor", $Q$ is then computed by comparing the submissions of participants to the true characteristics of the simulations, known only to the organizers of the challenge.

Each GREAT10 image set is designed to evaluate the ability of competing methods to deal with galaxy or PSF fields with different properties (e.g. constant versus varying field, size, SNR). We summarize in Table~\ref{table:property per set} the main PSF and galaxy characteristics for each of the GREAT10 data set.  

The exact metric used for assessing various characteristics of the methods have been described in the GREAT10 Galaxy challenge paper \citep{G10results}. It consists of:

\begin{itemize}
\item A ``raw''  quality factor $Q$ that measures the difference, averaged over all sets, between the reconstructed and true shear power spectra. The value of the raw $Q$ factor is used to rank the different submissions to the challenge and to display the leaderboard of the competition.
\item A quality factor $Q_{dn}$ is also obtained, after removal of biases caused by finite SNR or inherent shape measurement method-related noise.
\item A split of the total bias on the galaxy ellipticities into an additive and a multiplicative biases, respectively denoted as $c$ and $m$. 
\item Additional parameters $\mathcal{A}=\sigma^2(c)$ and $\mathcal{M} \simeq m^2 + 2m$, intended to measure the additive and multiplicative biases calculated at power spectrum level.
\end{itemize}

\begin{figure*}[t!]
\begin{center}
\includegraphics[scale=1.0]{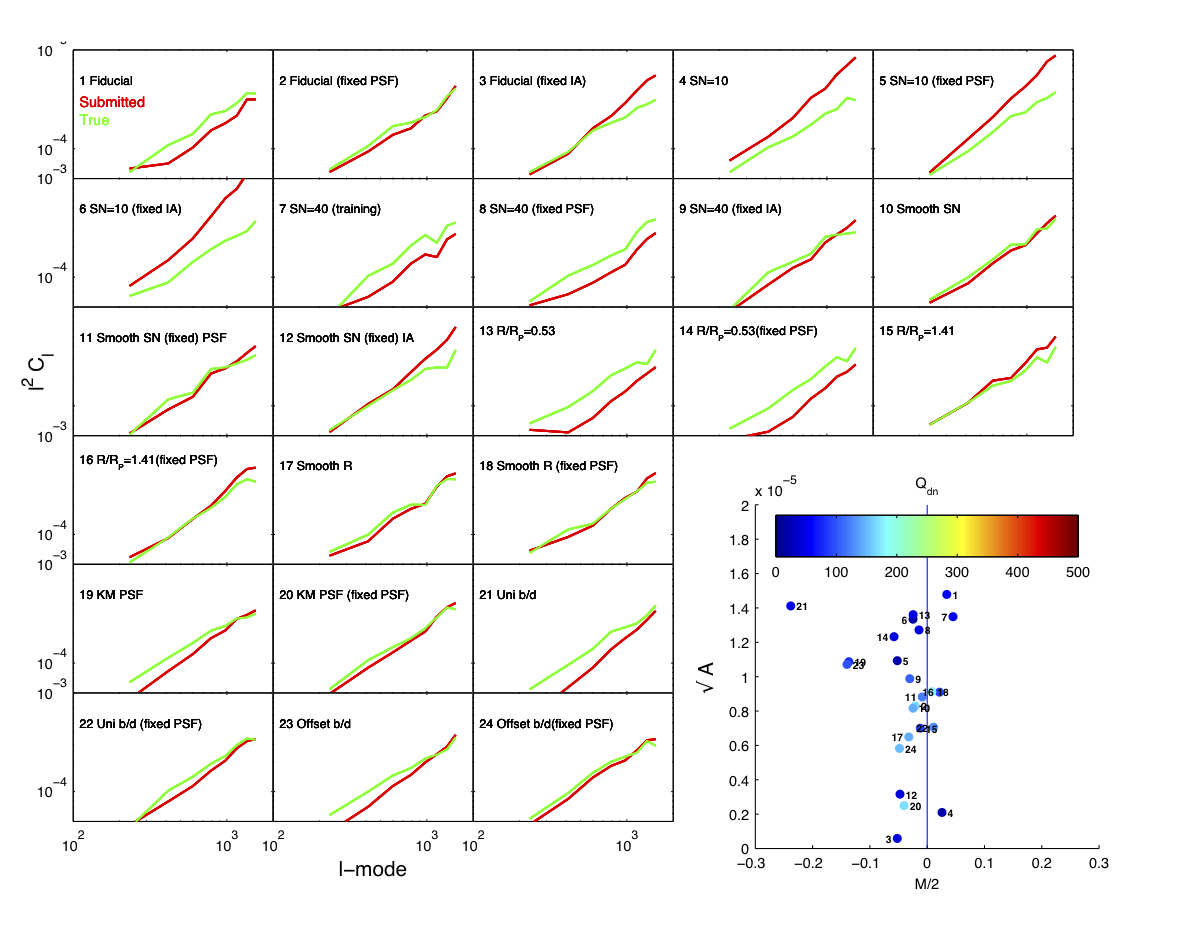}
\caption{Shear power spectra for the 24 data sets of the GREAT10 challenge, for our submission ``23/den/star" (see Table \ref{table:Q factor per method}). The red lines display our estimate of the shear power spectra while the green lines represent the true shear power.}
\label{fig:TVNN23_power_spectra}
\end{center}
\end{figure*}

\subsection{GREAT10 results}
\label{subsection:GREAT10 results}

While a general analysis of the GREAT10 results for all participating teams is available in \citet{G10results}, we focus here on a more specific analysis of the results obtained using our HNN deconvolution.  

In order to reduce the CPU time, we used smaller postage stamps of the galaxy images:  $19 \times 19$ , $23 \times 23$  and $29 \times 29$  pixels. Our submissions to GREAT10 were done under the names NN19 (official challenge submission) and NN23 (post-challenge submission) which stands for ``Neural Network", followed by the stamp size of the galaxy images, in pixels. In both submissions, we used the PSF under its functional form, but we reconstructed the PSF on a $15 \times 15$ postage stamp to carry out the deconvolution. 

We also made submissions after the deadline of the competition and in some cases, we applied image denoising to the data prior to the deconvolution, following \citet{Nurbaeva2011}. The submissions with denoising are labeled ``den" in Table~\ref{table:Q factor per method} and the ones without denoising are labeled ``raw". In these submissions we used  postage stamps with 23 pixels on-a-side and we used both the functional (labeled ``func") form for the PSF  and the pixelated FITS PSFs (labeled ``star"). 

\begin{figure*}[t!]
\begin{center}
\includegraphics[scale=0.45]{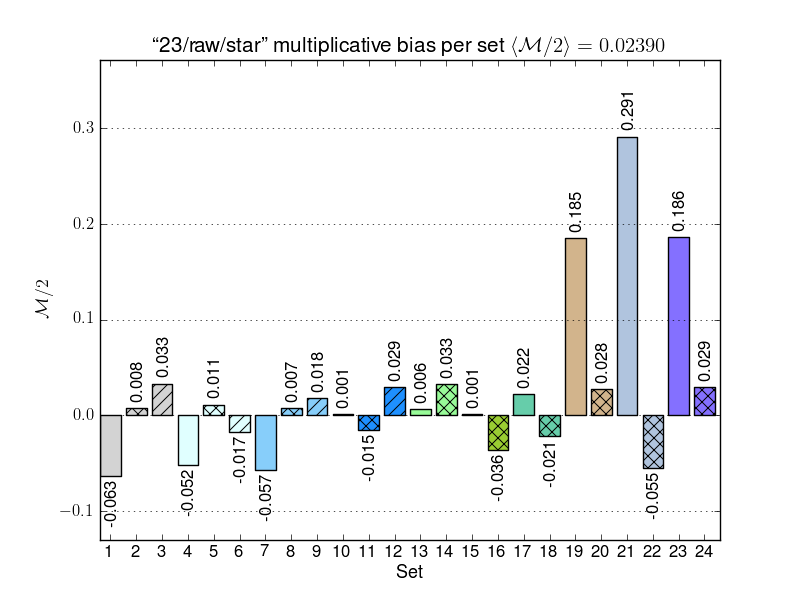}
\includegraphics[scale=0.45]{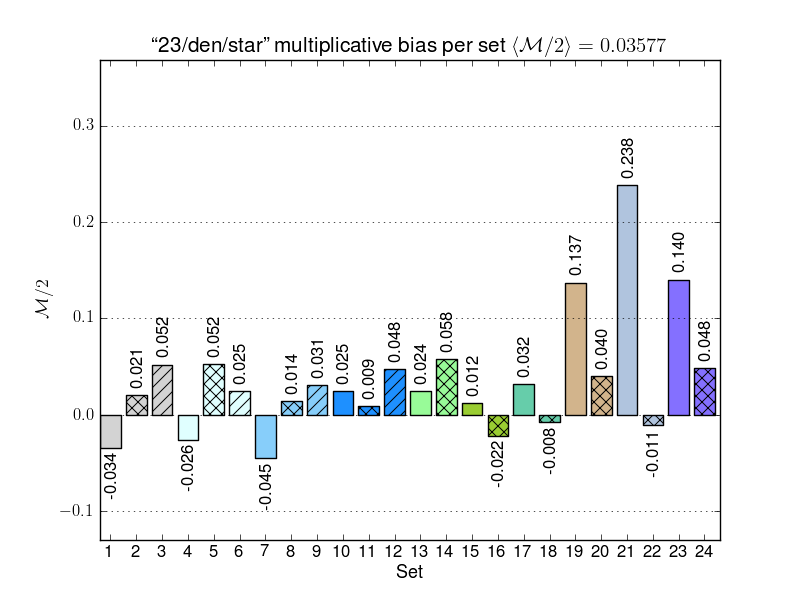}
\caption{Mutiliplicative bias per set, $\mathcal{M}/2$ on the shear power spectrum, as defined in \citet{G10results}. The $\mathcal{M}/2$ values are given on the left panel for our 23/raw/star submission and the right panel shows the 23/den/star submission.}
\label{fig:TVNN23_mult}
\end{center}
\end{figure*}
\begin{figure*}[t!]
\begin{center}
\includegraphics[scale=0.45]{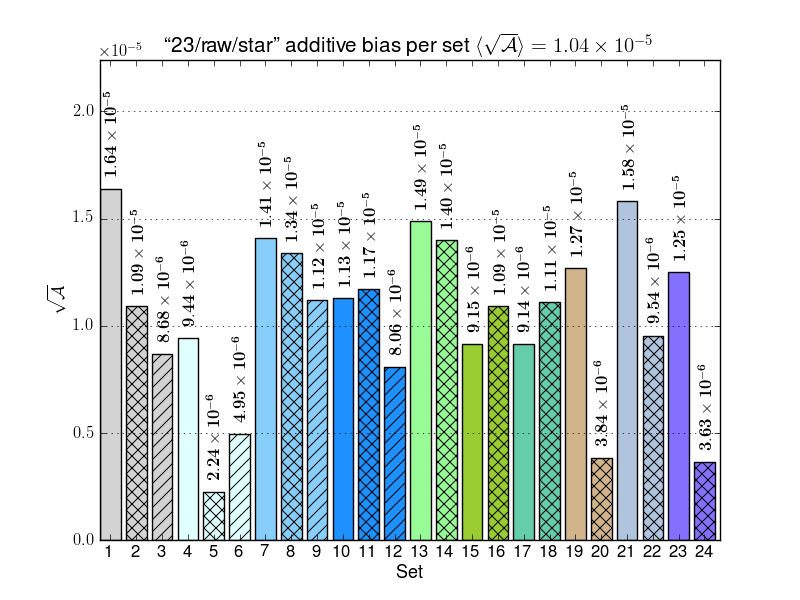}
\includegraphics[scale=0.45]{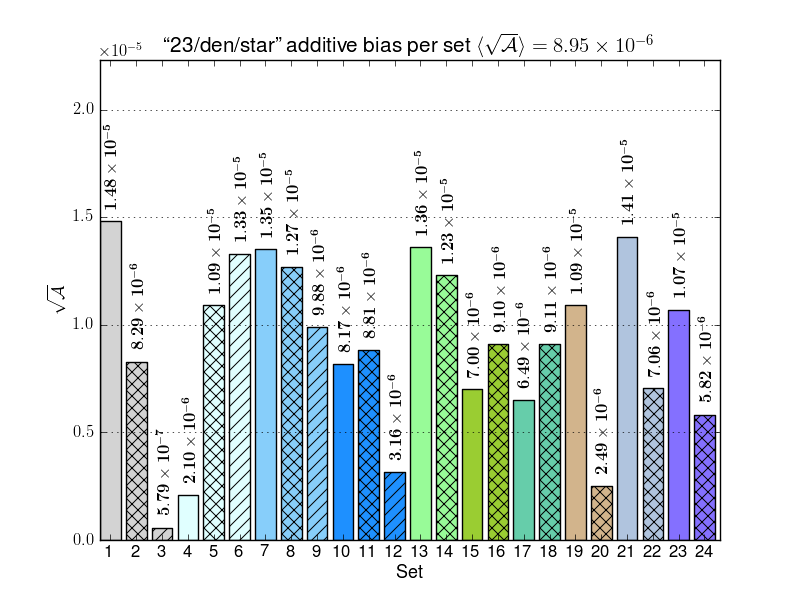}
\caption{Additive bias per set, $\sqrt{\mathcal{A}}$ on the shear power spectrum, as defined in \citet{G10results}. The $\sqrt{\mathcal{A}}$ values are given on the left panel for our 23/raw/star submission and the right panel shows the 23/den/star submission.}
\label{fig:TVNN23_add}
\end{center}
\end{figure*}

\subsection{Analysis of the GREAT10 results}
\label{subsection:analysis of results}

The quality factors, $Q$, obtained for our submissions are summarized in Table~\ref{table:Q factor per method} as well as in Fig.~\ref{fig:fig_Qcomparision}. Fig.~\ref{fig:TVNN23_power_spectra} shows the power spectra for the ``23/den/star" submission, i.e., for a deconvolution using $23 \times 23$ pixels stamps, denoising, and the pixelated version of the PSF. Finally, Figs.~\ref{fig:TVNN23_mult} \& \ref{fig:TVNN23_add} show the multiplicative bias $\mathcal{M}/2$, and additive bias $\sqrt{\mathcal{A}}$ computed for the shear power spectra.  

It is interesting to analyze first the influence of our different submissions on the $Q$ factors displayed in Fig.~\ref{fig:fig_Qcomparision}. Before doing so, we should note that our first submission, NN19, was affected by a misunderstanding of the centering convention in the GREAT10 challenge. This resulted in a very low $Q$ factor for this
submission, which we discard in the following discussion. In the other submissions, we note that increasing the stamp
size used for the deconvolution significantly improves the $Q$ factor. A stamp size of 29 pixels, however, results in large cpu times, of the order of 4 sec. per galaxy. This would make the method very hard to use in practice on data from a real survey.

All of our mean $Q$ factors are between 70 and 90, meaning that on average the use of the functional form for the PSF or the pixelated PSF has little influence. This may be due to the fact that the signal-to-noise un the pixelated PSFs is high enough to make the functional form unnecessary. We also note that denoising improves the $Q$ factor by 5 to 10\%. 


One can see that the good $Q$ factors are obtained for the sets 10 and 11, which have a realistic signal-to-noise distribution. The same is observed for Sets 17 \&18 which contain smooth galaxies. Among these 4 sets, we see no difference in terms of $Q$ when the PSF is kept fixed or variable. 

Finally, maybe the most interesting observations are for Set 20 which contains turbulent PSFs and for Sets 23 \& 24, which contain offset bulge/disk galaxies. Both characteristics are typical for realistic ground-based surveys which will contain complex PSFs and galaxies. Obtaining our best $Q$ factors for these sets is therefore very encouraging. The purpose of direct image deconvolution is indeed to enable the use of any arbitrary galaxy shape, independent of a simplistic underlying model. This gain in galaxy complexity is made to the prize of a lower $Q$ factor for sets with low signal-to-noise galaxies and for small galaxies. This is not unexpected as deconvolution in general tend to enhance noise and to perform less well as sampling is getting worse, i.e., exactly for small and faint galaxies. 

Obtaining good $Q$ factors is necessary but not sufficient to allow weak lensing measurements. We also need to ensure that the shape measurement method does not introduce significant bias on the shear power spectrum. These biases are summarised in Figs.~\ref{fig:TVNN23_power_spectra}-\ref{fig:TVNN23_add}. It is interesting to see that, while the $Q$ factors for the low signal-to-noise galaxies and small galaxies are poor, the biases on the shear power spectrum remain comparable with those of sets with high signal-to-noise and large galaxies. In addition, the biases on the offset bulge-disk galaxies remain low. Finally, we see that denoising decreases the multiplicative bias and slightly increases the additive bias and this almost independently of the properties of the galaxies and PSFs in each set. 

\section{Conclusion}
\label{section:conclusion}

We have introduced a new method to spatially deconvolve images, and we have applied it 
for the first time to the field of weak gravitational lensing. While the deconvolution process tends to enhance the noise, it has the advantage to allow ellipticity measurements with no assumption of an underlying light profile. The noise enhancement itself is in part circumvented by measuring the galaxy ellipticities on the autocorrelation function of the deconvolved image.  Using the autocorrelation function, has the  advantage to be always centered the array of pixels, performing the higher accuracy of the shape measurement. 

We have then confronted our algorithm to the simulated data proposed in the GREAT10 image processing challenge. 
We find that image deconvolution performs well on complex galaxy shapes although with somewhat lower performances at low signal-to-noise and for small galaxies (compared with the PSF size). These results are almost independent on the PSF properties, i.e., shape and spatial variations. We also find that image denoising improves the precision of the measurement ($Q$ factor) without affecting much the accuracy, measured by the multiplicative and additive biases on the shear power spectrum. 

The above behaviour of our method is observed for the specific data of the GREAT10 challenge and would need more investigations either on more complex simulations or on real data. The method remains cpu-intensive and requires in practice about 1 s per galaxy to converge properly. The results obtained on GREAT10 are nevertheless encouraging, as the method performs best on more complex galaxy shapes. Application to the future data of the GREAT3 challenge (\citet{2013GREAT3}), which will include HST-like galaxies, will be a crucial test for direct image deconvolution. 

\begin{acknowledgements}
 We are grateful to the GREAT10 Coordination Team for organizing this stimulating challenge, in particular Tom Kitching for his help and for sharing the shear analysis code. We thank  Matthew Luke Nichols for helpful discussions. Malte Tewes acknowledges support by the DFG grant Hi 1495/2-1. This work is supported by the Swiss National Science Foundation (SNSF).  The GREAT10 challenge itself was sponsored by a EU FP7 PASCAL 2 challenge grant.  
 \end{acknowledgements}

\bibliographystyle{aa}
\bibliography{ArticlesTVNN}

\end{document}